\documentclass[aps,preprint,showpacs,showkeys,superscriptaddress]{revtex4-1}

\usepackage{dcolumn}
\usepackage{bm}
\usepackage{amssymb}
\usepackage{amsfonts}
\usepackage{amsmath}
\usepackage{graphicx}
\usepackage{epstopdf}
\usepackage{float}
\usepackage{relsize}
\usepackage{mathtools}
\usepackage{placeins}
\usepackage{array}
\usepackage{xcolor}
\begin{document}

\title{Kiselev black holes in $f(R,T)$ gravity}

\author{L. C. N. Santos}
\email{luis.santos@ufsc.br}

\affiliation{Departamento de Física, CCEN--Universidade Federal da Paraíba; C.P. 5008, CEP  58.051-970, João Pessoa, PB, Brazil} 

\author{F. M. da Silva}
\email{franmdasilva@gmail.com}

\affiliation{Núcleo Cosmo--ufes \& Departamento de Física, Universidade Federal do Espírito Santo, Av. Fernando Ferrari, 540, CEP 29.075-910, Vitória, ES, Brazil}

\author{C. E. Mota}
\email{clesio200915@hotmail.com}

\affiliation{Departamento de Física, CFM--Universidade Federal de Santa Catarina; C.P. 476, CEP 88.040-900, Florianópolis, SC, Brazil} 

\author{I. P. Lobo}
\email{lobofisica@gmail.com}

\affiliation{Department of Chemistry and Physics, Federal University of Para\'iba, Rodovia BR 079 - Km 12, 58397-000 Areia-PB,  Brazil.}

\affiliation{Physics Department, Federal University of Lavras, Caixa Postal 3037, 37200-000 Lavras-MG, Brazil.}

\author{V. B. Bezerra}
\email{valdir@fisica.ufpb.br}

\affiliation{Departamento de Física, CCEN--Universidade Federal da Paraíba; C.P. 5008, CEP  58.051-970, João Pessoa, PB, Brazil}

\begin{abstract}
We obtain new exact solutions for the gravitational field equations in the context of $f(R,T)$ gravity, thereby obtaining different classes of black holes surrounded by fluids, taking into account some specific values of the parameter of the equations of state, $w$. In order to obtain these solutions in the context of $f(R,T)$ gravity, we consider viable particular choices of the $f(R,T)$.
Considering an anisotropic energy-momentum tensor, we write the field equations with the required symmetries for this type of solution. Then, we analyze the conditions of energy in a general way and also for particular values of the parameter $w$ of the equation of state. In addition, thermodynamic quantities, such as Hawking temperature and mass associated to the horizons of solutions, are taken into account in our analysis.

\end{abstract}


\maketitle

\preprint{}

\volumeyear{} \volumenumber{} \issuenumber{} \eid{identifier} \startpage{1} %
\endpage{}
\section{Introduction}
The observations showing the accelerating expansion of the universe \cite{supernova} are one of the most important discoveries in cosmology in recent years. In order to explain such accelerating expansion, several models have been studied in general relativity and modified theories of gravity. In the context of general relativity, we can assume equations of state connecting the energy density and the pressure associated to the universe, which demand a negative pressure for the equations of state. Assuming a relation in the form $p=w\rho$, where $p$ is the pressure and $\rho$ the energy density, 
the physical properties of the matter and energy of space-time depend on the values of the parameter $w$. 

Kiselev proposed a general relation connecting energy density and pressure \cite{kiselev}, where the components of the energy-momentum tensor can be associated to an anisotropic fluid that by taking the isotropic average over the angles, one obtains 
the barotropic equation of state. In this way, particular choices of the parameter of equations of state can be assumed in this formulation and some of these values reproduce, in the cosmological context, the accelerating pattern \cite{kiselev}. Kiselev solution \cite{kiselev} was also associated to a quintessence field, (See \cite{matt} for a discussion on the terminology issues concerning this solution), and, in this context, several investigations have been done on the shadow of black holes \cite{shadow1,shadow2,shadow3,shadow4,shadow5}, quasinormal modes
\cite{quasi1,quasi2,quasi3,quasi4,quasi5,quasi6,quasi7,quasi8,quasi9,quasi10,quasi11},  thermodynamics of black holes\cite{termo1,termo2,termo3,termo4,termo5,termo6,termo7,termo8,termo9,termo10,termo11,termo12,termo13,termo14,termo15}, and other questions related to black holes surrounded by fluids in the framework of nonconservative gravity \cite{rastallbh1,rastallbh2,rastallbh3,rastallbh4,rastallbh5,rastallbh6}. In which concerns the singularities present in several classes of black holes, the weak cosmic censorship conjecture demands that no naked singularities exist in the space-time, i.e, the singularities arising from the solutions of Einstein field equations are hidden within event horizons \cite{penrose}. Although this conjecture is violated in solutions such as the extreme Reissner-Nordstr\"{o}m and extreme Kerr black hole,  some studies \cite{censura} reveal that it can be satisfied in Reissner-Nordström-AdS black hole surrounded by quintessence. 

On the other hand, some extensions of the general relativity have gained interest in recent days due to the possibility of their use for addressing open problems in both astrophysical \cite{nashed2020nontrivial,black2,black3,santos00,retro27,santos8,rainbow6} and cosmological context \cite{carroll2004,starobinsky,barrow1983}. In particular, motivated by the idea of considering different classes of coupling between matter and geometry in a general formalism, and by advances in cosmology due to the $f(R)$ theories, it was proposed the $f(R,T)$ gravity theory \cite{fdr2}. It is well-known that theories with nontrivial matter-geometry coupling have additional effects on the dynamics of the bodies in the space-time. Indeed, motion of the massive particles in $f(R,T)$ is non-geodesic and undergoes an extra force depending on the coupling of the matter and the geometry. The possibility of reconstructing arbitrary Friedmann-Robertson-Walker cosmologies by an appropriate choice
of a function $f(T)$ has stimulated the study of the $f(R,T)$ in cosmological scenarios \cite{cosmo1,cosmo2,cosmo3}. 

Such effects may become more prominent for high densities and pressures, in this way, it is natural to test the effects of modification of gravity imposed by $f(R,T)$ theory in the scale of compact objects. The influence of the dependency on the Ricci scalar $R$ considering vacuum solutions has been studied in the literature in the context of $f(R)$ gravity. In the case of theories of gravity in which the Lagrangian density depends on $T$,  it is expected differences between solutions in these models and general relativity in the presence of a no-zero energy-momentum tensor, i.e, it is expected additional effects due to the matter-geometry coupling. In particular, the presence of fluids surrounding spherical sources of matter can be an interesting system to study the effects of this coupling between matter and geometry. In this paper, we study nontrivial black hole solutions for the field equations in the context of $f(R,T)$ gravity where the black hole is surrounded by the fluid discussed by Kiselev \cite{kiselev}. In addition, it is considered particular cases associated to the solution obtained by taking into account the appropriate values of the parameter of fluid equation of state. 
It is worth emphasizing that we will consider viable particular choices of the $f(R,T)$ function in such a way that the obtained results can be associated to an extension of the Kiselev solution \cite{kiselev} for this modified theory of gravity.

The paper is organized in the following way: In section II, we review the field equations in the context of $f(R,T)$ gravity and particularize it for a specific choice of $f(R,T)$ function. Section III is dedicated to the study of energy-momentum tensor associated to the fluid surrounding the black hole. In section IV, we discuss the $f(R,T)$ gravity model in which  the trace function is written in terms of an arbitrary exponent and we obtain an exact solution corresponding to a black hole surrounded by fluids. We also discuss general energy conditions, horizons, mass, and temperature for this solution.  The analysis of the cases corresponding to particular
choices of the parameter of the equation of state $w$ and their relation with
 solutions in the context of general relativity is considered in section V. Finally, in section VI, we present our final remarks.

\section{Field equations}
In this section, we briefly review the field equations in the context of $f(R,T)$ gravity \cite{fdr2}. In this formulation, it is assumed an action in the form
\begin{equation}
S=\frac{1}{16\pi}\int f(R,T)\sqrt{-g}d^4x + \int L_{m}\sqrt{-g}d^4x,
    \label{e1}
\end{equation}
with $f(R,T)$ being a function of the Ricci scalar, $R$, and of the trace $T$ of the  energy-momentum tensor of the
matter. Note that $L_m$ in Eq. (\ref{e1}) represents the matter Lagrangian density, this term is associated to a particular  energy-momentum tensor. By varying the action $S$ with respect to  the metric tensor, we obtain the integral

\begin{align}
\delta S=& \frac{1}{16\pi}\int\left[f_{R}(R,T)R_{\mu\nu}\delta g^{\mu\nu} + f_{R}(R,T)g_{\mu\nu} \Box \delta  g^{\mu\nu}  \right. +  
\\
& -f_{R}(R,T)\nabla_{\mu}\nabla_{\nu}\delta g^{\mu\nu} + 
 f_{T}(R,T)\frac{\delta (g^{\eta\xi}T_{\eta\xi})}{\delta g^{\mu\nu}}\delta g^{\mu\nu}+
 \\
 & \left. -\frac{1}{2}g_{\mu\nu}f(R,T)\delta g^{\mu\nu}+\frac{16\pi}{\sqrt{-g}}\frac{\delta (\sqrt{-g}L_{m})}{\delta g^{\mu\nu}}\right]\sqrt{-g}d^4x  
    \label{e2}
\end{align}
where $f_R(R,T)=\partial f(R,T)/\partial R$ and $f_T(R,T)=\partial f(R,T)/\partial T$. By integrating the second and third terms and considering the variation of $T$ as 
\begin{equation}
\frac{\delta (g^{\eta\xi}T_{\eta\xi})}{\delta g^{\mu\nu}}=T_{\mu\nu} + \Theta_{\mu\nu}
    \label{e3},
\end{equation}
where
\begin{equation}
\Theta_{\mu\nu} \equiv g^{\eta\xi}\frac{\delta T_{\eta\xi}}{\delta g^{\mu\nu}}=-2T_{\mu\nu}+g_{\mu\nu}L_{m}-2g^{\eta\xi}\frac{\partial^2L_{m}}{\partial g^{\mu\nu}g^{\eta\xi}},
    \label{e4}
\end{equation}
emerges from the definition of the variation of the matter Lagrangian, we obtain 
\begin{align}
 f_{R}&(R,T)R_{\mu\nu}-\frac{g_{\mu\nu}}{2}f(R,T)+
(g_{\mu\nu}\Box - \nabla_{\mu}\nabla_{\nu})f_{R}(R,T) \nonumber\\
&= 8\pi T_{\mu\nu} - f_{T}(R,T)T_{\mu\nu}- f_{T}(R,T)\Theta_{\mu\nu}\, ,
   \label{e5}
\end{align}
 which is the form of the field equation in $f(R,T)$ gravity that we are going to use in this paper. In the special case in which $f(R,T) \equiv f(R)$, Eq. (\ref{e5}) reduces to the field equations in the context of $f(R)$ gravity. In this way, the novel feature introduced by $f(R,T)$ gravity is the possibility of arbitrary coupling between matter
and geometry. An interesting choice of $f(R,T)$ functions is given by \cite{fdr2}
\begin{equation}
    f(R,T)=R + 2f(T),
    \label{e6}
\end{equation}
where $f(T)$ is an arbitrary function of the  trace of the energy-momentum tensor. From Eq. (\ref{e5}) and by considering the trace function Eq. (\ref{e6}), the field equations are given by 
\begin{align}
 R_{\mu\nu}-\frac{g_{\mu\nu}}{2}R=& 8\pi T_{\mu\nu}- 2f'(T)T_{\mu\nu}
 \nonumber\\
 &- 2f_{T}'(T)\Theta_{\mu\nu} + f(T)g_{\mu\nu},
    \label{e7}
\end{align}
where $f'(T)=df(T)/dT$. Concerning the choice of the function $f(T)$, before we write an explicit expression, it is instructive to discuss the form of the trace associated to a specific choice of the energy-momentum tensor of the matter. As we will see, the condition
of additivity and linearity imposed to the Kiselev solution naturally restricts the form of this function. 

\section{Energy-momentum of the Kiselev black hole}
Considering a spherically symmetric space-time, the line element associated to a static geometry can be written as 
\begin{equation}
    ds^2=B(r)dr^2-A(r)dr^2-r^2(d\theta^2+\sin\theta^2d\phi^2),
    \label{metric}
\end{equation}
where $B(r)$ and $A(r)$ are unknown functions of the coordinate $r$. 
The energy-momentum tensor in Kiselev black holes is defined to have the components of the spatial sector proportional to the time sector:
\begin{align}
    T^{t}_{\:\:\:t}= T^{r}_{\:\:\:r}&=\rho(r), \label{e8}\\
   T^{\theta}_{\:\:\:\theta}= T^{\phi}_{\:\:\:\phi}&=-\frac{1}{2}\rho (3w+1),
   \label{e9}
\end{align}
and $w$ is the parameter of equation of state.
In addition, by taking the isotropic average over the angles, in the place of equations (\ref{e8}) and (\ref{e9}), one obtain the barotropic equation of state $p=w\rho$. Kiselev black\cite{kiselev}
holes have the components of  energy-momentum tensor effectively connected to an anisotropic fluid  represented by 
\begin{equation}
    T^{\mu}_{\:\:\:\nu} = diag(\rho,-p_r,-p_t,-p_t),
    \label{e10}
\end{equation}
where $p_r=-\rho$ and $p_t=\frac{1}{2}\rho (3w+1)$,
 which can be extracted from the general form of the anisotropic fluid  \cite{santos00}:
\begin{equation}
    T_{\mu\nu}=-p_{t}g_{\mu\nu}+(p_{t}+\rho)U_{\mu}U_{\nu}+(p_{r}-p_{t})N_{\mu}N_{\nu}
    \label{anytensor},
\end{equation}
where $p_{t}(r)$, $\rho(r)$ and $p_{r}(r)$ are  the tangential or transverse pressure, the energy density and the radial pressure of the fluid, respectively. The quantities $U_{\mu}$ and $N_{\mu}$ represent the four velocity and radial unit vector, respectively, are defined as
 \begin{equation}
        U^{\mu}=\left( \frac{1}{\sqrt{B(r)}},0,0,0\right)
    \label{eq1},
 \end{equation}
 \begin{equation}
        N^{\mu}=\left( 0,\frac{1}{\sqrt{A(r)}},0,0\right)
    \label{eq2},
 \end{equation}
and obey the conditions $U_{\nu}U^{\nu}=1$, $N_{\nu}N^{\nu}=-1$ and $U_{\nu}N^{\nu}=0$. The matter Lagrangian density associated to the anisotropic fluid is given by $L_{m}=(-1/3)(p_r+2p_t)$ \cite{lm}. This implies that Eq. (\ref{e4}) can be written as
\begin{equation}
    \Theta_{\mu\nu}=-2T_{\mu\nu}-\frac{1}{3}(p_r+2p_t)g_{\mu\nu}.
\end{equation}
In the next section, we use these results in the context of the Kiselev solutions of black holes \cite{kiselev} that demands additivity and linearity  between
the metric components. 

\section{The $f(T)=\varkappa T^{n}$ model}
 
In this section, we consider the case where the trace is written in terms of an arbitrary exponent. Substituting the energy-momentum tensor (\ref{e10}) and $f(T)=\varkappa T^{n}$ into the field equation (\ref{e7}), leads to
\begin{align}
    G^{t}_{\:\:\:t}&=H(\rho) \label{eq3},\\
    G^{r}_{\:\:\:r}&=H(\rho)\label{eq4}, \\
      G^{\theta}_{\:\:\:\theta}&=F(\rho) \label{eq5},
\end{align}
where the functions are given by
\begin{align}
    H(\rho)=& 8\pi\rho-\frac{2n(w+1)(\varkappa\rho-3\varkappa\rho w)^n}{3w-1}+(\varkappa\rho-3\varkappa\rho w)^n, \label{e10a}\\
    F(\rho)=& -(12w+4)\pi\rho+\frac{n(w+1)(\varkappa\rho-3\varkappa\rho w)^n}{3w-1}+\nonumber\\
    &+(\varkappa\rho-3\varkappa\rho w)^n,
    \label{e11}
\end{align}
and were obtained from
\begin{equation}
    G^{\mu}_{\:\:\:\nu}=R^{\mu}_{\:\:\:\nu}-\frac{1}{2}\delta^{\mu}_{\:\:\:\nu}R.
    \label{e12}
\end{equation}
Equations (\ref{eq3}),(\ref{eq4}) and (\ref{eq5}) form the independent set of field equations for black holes surrounded by a fluid whose components of the energy-momentum tensor are given by Eqs.(\ref{e8}) and (\ref{e9}). As required in Kiselev approach\cite{kiselev}, Eqs. (\ref{eq3}) and (\ref{eq4}) yield the relation 
\begin{equation}
    G^{t}_{\:\:t}=G^{r}_{\:\:r}
     \label{e13}.
\end{equation}
The symmetry arising from Eq. (\ref{e13}) demands that 
\begin{equation}
    B(r)\frac{dA(r)}{dr}+A(r)\frac{dB(r)}{dr}=0
    \label{eq8},
\end{equation}
and, as a consequence $A(r)=1/B(r)$. Substituting this result in the original set of field equations, we obtain 
\begin{align}
   G^{t}_{\:\:t}=G^{r}_{\:\:r}=& -\frac{1}{r}\frac{dB(r)}{dr} - \frac{B(r)}{r^2} + \frac{1}{r^2}= H(\rho), \label{eq10a}\\
 G^{\theta}_{\:\:\theta}=G^{\phi}_{\:\:\phi}=& -\frac{1}{2}\frac{d^2B(r)}{dr^2}-\frac{1}{r}\frac{dB(r)}{dr} = F(\rho)
    \label{eq10}.
\end{align}
Now, the conditions supposed in \cite{kiselev}, including additivity and linearity, restricts the form of functions $H$ and $F$. If one assumes a relation $H=kF$, where $k$ is an arbitrary constant, then the exponent of the function $f(T)$ must be $n=1$.  
Combining equations (\ref{eq10a}) and (\ref{eq10}) with the aforementioned assumptions, the energy density dependence is eliminated, and we obtain the following result
\begin{align}
\frac{3\varkappa+8\pi-w\varkappa}{4(3\pi w + w\varkappa+\pi)}\left(\frac{1}{2}\frac{d^2B(r)}{dr^2}+\frac{1}{r}\frac{dB(r)}{dr}\right) + & \nonumber\\
 + \frac{1}{r}\frac{dB(r)}{dr} + \frac{B(r)}{r^2} - \frac{1}{r^2}=0&. 
    \label{eq11}
\end{align}
Equation (\ref{eq11}) can be rewritten in the following way
\begin{equation}
    \begin{split}
        & \frac{1}{r^2} \left( \frac{d}{dr} (r B(r)) -1 \right) =  \\ 
        & -\frac{1}{2 r} \left(\frac{3\varkappa+8\pi-w\varkappa}{4(3\pi w + w\varkappa+\pi)}\right) \frac{d }{dr} \left( \frac{d}{dr} (r B(r)) -1 \right),
    \end{split}
    \label{eq12}
\end{equation}
which can be easily integrated in order to get: 
\begin{equation}
    \frac{d}{dr} (r B(r)) -1  = cr^{-\frac{8(3\pi w + w\varkappa+\pi)}{3\varkappa+8\pi-w\varkappa}},
    \label{eq13}
\end{equation}
with $c$ being an integration constant. By integrating once again, we get 
\begin{equation}
    B(r)=1+\frac{c_{1}}{r}+Kr^{-\frac{8(3\pi w + w\varkappa+\pi)}{3\varkappa+8\pi-w\varkappa}},
    \label{eq14}
\end{equation}
where $c_{1}$ and $K$ are constants. Thus, substituting (\ref{eq14}) in the field equation, we obtain the following expression for the energy density
\begin{equation}
   \rho= D{r}^{-6\,{\frac { \left( w+1 \right)  \left( 4\,\pi +{\it \varkappa} \right) }{-w{\it \varkappa}+8\,\pi +3\,{\it \varkappa}}}},
   \label{14}
\end{equation}
where
\begin{equation}
D \equiv 3{\frac {K \left( 8\,\pi \,w+3\,w{\it \varkappa}-{\it \varkappa} \right) }{{w}^{2}{{\it \varkappa}}^{2}-16\,\pi \,w{\it \varkappa}\\
\mbox{}-6\,w{{\it \varkappa}}^{2}+64\,{\pi }^{2}+48\,\pi \,{\it \varkappa}+9\,{{\it \varkappa}}^{2}}}.
    \label{15}
\end{equation}
 By identifying $c_{1}=-2M$, where $M$ is the total mass of the black hole surrounded by the fluid, the solution obtained in this work reduces to the Schwarzschild solution in the absence of fluid.

\subsection{Energy conditions}
\begin{figure}[H]
\centering
\includegraphics[scale=0.27]{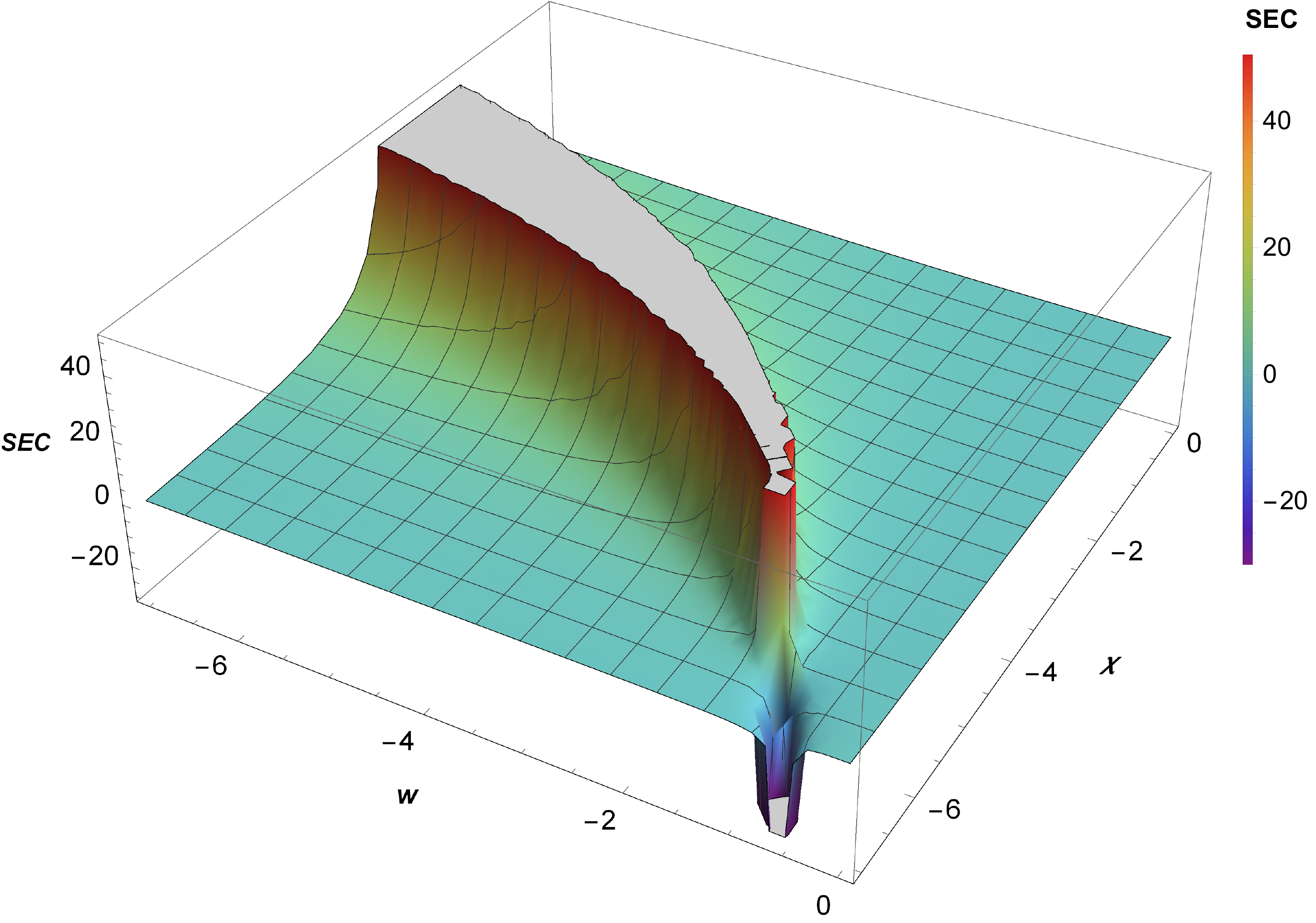} 
\caption{Condition (\ref{e19a}) is plotted as a function of $w$ and $\varkappa$ considering positive values of $K$. Negative values of the independent variable are associated to the regions where the SEC is violated.   
}
\label{fig1}
\end{figure}

\begin{figure}[h]
\centering
\includegraphics[scale=0.27]{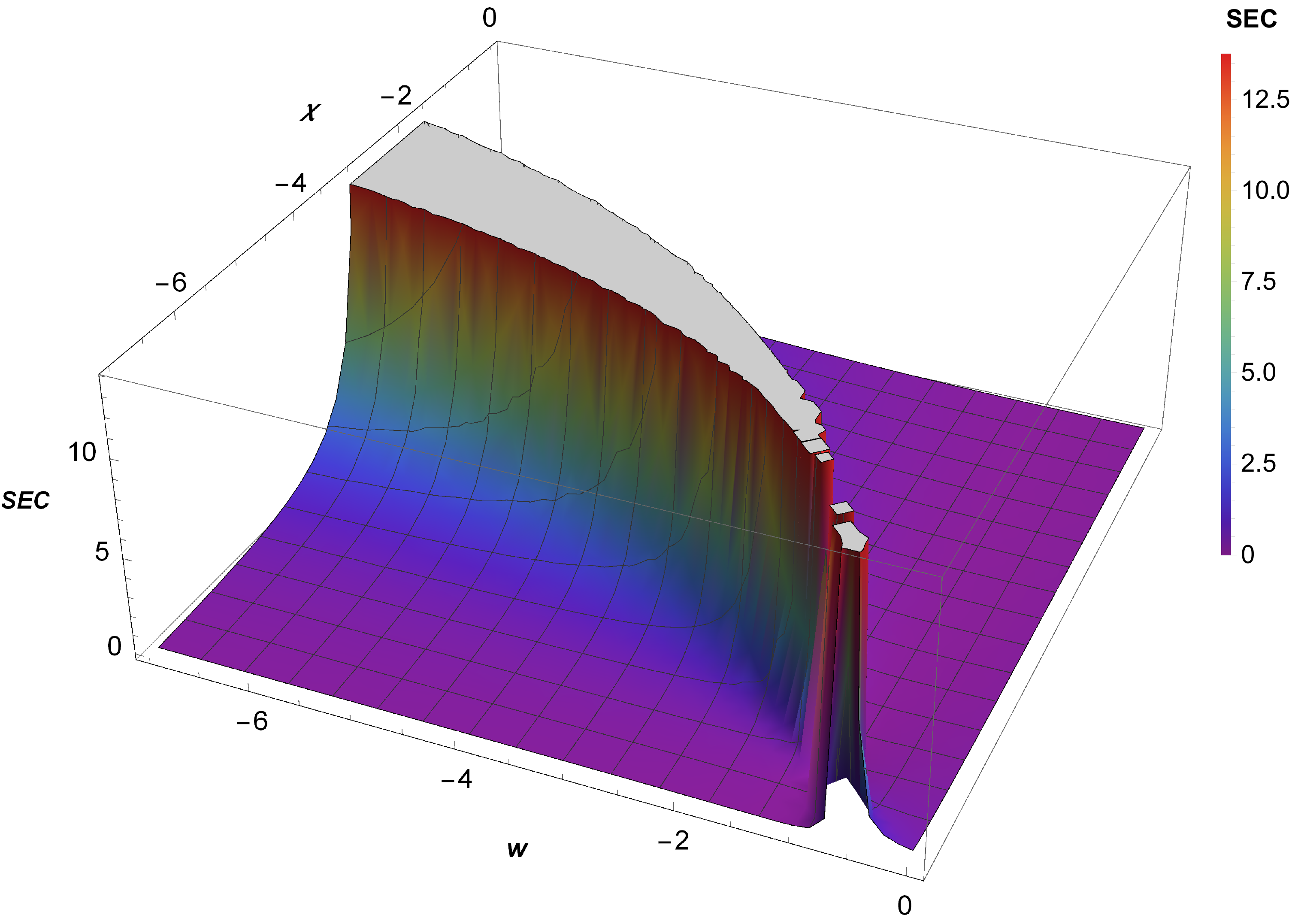} 
\caption{Condition (\ref{e19b}) is plotted as a function of $w$ and $\varkappa$ considering positive values of $K$. Negative values are associated to the regions where the SEC is violated.   
}
\label{fig2}
\end{figure}

Regarding the energy conditions for the anisotropic fluid, there are some requirements that the components of energy-momentum must satisfy in order to represent a realistic matter distribution.  It is well-known that exotic matter violate certain energy conditions of the energy-momentum tensor. In the case of the strong energy condition (SEC), the expression for anisotropic fluids is given by the pair of equations 
\begin{equation}
\text{SEC}:\:\: \rho + p_{n} \geq 0, \:\:\: \rho + \sum_{n} p_{n} \geq 0,
    \label{eq16}
\end{equation}
where $n=1,2,3...$  By considering the energy density (\ref{14}) and the components of pressure defined in (\ref{e10}), we obtain the radial and tangential pressure in the form 
\begin{align}
p_r=&-\rho=-D{r}^{-6\,{\frac { \left( w+1 \right)  \left( 4\,\pi +{\it \varkappa} \right) }{-w{\it \varkappa}+8\,\pi +3\,{\it \varkappa}}}},
    \label{e14}\\
p_t=&\frac{1}{2}(3w+1)D{r}^{-6\,{\frac { \left( w+1 \right)  \left( 4\,\pi +{\it \varkappa} \right) }{-w{\it \varkappa}+8\,\pi +3\,{\it \varkappa}}}}.
\label{e15}
\end{align}
In this way, the SEC  can be written as 
\begin{align}
    \rho+p_r=&0, \label{e16}\\    
    \rho+p_t=&\frac{3}{2}(w+1)D{r}^{-6\,{\frac { \left( w+1 \right)  \left( 4\,\pi +{\it \varkappa} \right) }{-w{\it \varkappa}+8\,\pi +3\,{\it \varkappa}}}},\label{e17}\\
    \rho+p_r+2p_t=&(3w+1)D{r}^{-6\,{\frac { \left( w+1 \right)  \left( 4\,\pi +{\it \varkappa} \right) }{-w{\it \varkappa}+8\,\pi +3\,{\it \varkappa}}}}.
    \label{e18}
\end{align}
As a consequence of Eqs. (\ref{e16}),(\ref{e17}) and (\ref{e18}), the conditions in which the SEC is satisfied are the following
\begin{align}
    \frac{K(8\pi w +3w\varkappa-\varkappa)(3w+1)}{w^2\varkappa^2-16\pi w \varkappa -6w \varkappa^2 +64\pi^2+48\pi\varkappa+9\varkappa^2} & \geq 0,\label{e19a}\\
    \frac{K(8\pi w +3w\varkappa-\varkappa)(w+1)}{w^2\varkappa^2-16\pi w \varkappa -6w \varkappa^2 +64\pi^2+48\pi\varkappa+9\varkappa^2} & \geq 0.
    \label{e19b}
\end{align}
Possible solutions that satisfy the above equations connecting the parameters $w$, $\varkappa$ and the SEC can be visualized in Fig. \ref{fig1} and Fig. \ref{fig2}. In Fig. \ref{fig1}, we plot the left-hand side (LHS) of Eq. (\ref{e19a}) and we consider positive values of $K$. In the case of Fig. \ref{fig2}, we plot the LHS of Eq. (\ref{e19b}) and also consider positive values of $K$.
In these plots, negative values of the independent variable (vertical axis) correspond to the region where the SEC is violated. Note that negative values of $K$ cause a reflection on the values of the independent variable. We can see that in both the plots, positive values of $K$ tend to produce results that do not violate the SEC. 

\subsection{Horizons, mass and temperature}
By denoting the place where the metric function $B(r)$ is equal to zero as  $r_h$, the horizons of a metric function are defined as $B(r_h)=0$. It follows from this definition that the black hole mass can be written in terms of $r_h$, as follows
\begin{equation}
M(r_h)=\frac{1}{2}r_h\left[1+K r_h^{-\frac{8(3\pi w + w\varkappa+\pi)}{3\varkappa+8\pi-w\varkappa}}\right],
    \label{t1}
\end{equation}
which represents a general relation connecting the mass, the parameters $w$ and $\varkappa$. The surface gravity of
the black hole surround by a fluid, given by $\kappa=\frac{1}{2}\frac{dB(r)}{dr}\Big|_{r=r_h},$ can be evaluated, with the following result
\begin{equation}
\kappa=\frac{8\pi + 3\varkappa-w\varkappa-3K(8\pi w-\varkappa + 3w\varkappa) r_h^{-\frac{8(3\pi w + w\varkappa+\pi)}{3\varkappa+8\pi-w\varkappa}}}{2(8\pi +3\varkappa - w\varkappa)r_h}.
    \label{t2}
\end{equation}
This quantity, in terms of the parameter of the equation of state $w$, and of the parameter of the $f(R,T)$ gravity $\varkappa$, gives us the Hawking temperature $T=\hbar \kappa/(2\pi $), which follows
\begin{equation}
T_{BH}=\frac{\hbar}{4\pi r_h}-\frac{ 3K \hbar (8\pi w-\varkappa + 3w\varkappa) r_h^{-\frac{8(3\pi w + w\varkappa+\pi)}{3\varkappa+8\pi-w\varkappa}}}{4\pi(8\pi +3\varkappa - w\varkappa)r_h}
    \label{t3}.
\end{equation}
From this expression,  we observe that Hawking temperature for a black hole surrounded by fluid in $f(R,T)$ gravity has an additional structure that comes from the dependence on the parameter $\varkappa$. In the following, we analyze the influence of this result by taking into account particular choices of the parameter $w$.

\section{Particular cases}
Let us study some cases corresponding to particular choices of the parameter of the equation of state $w$ and compare with the corresponding solutions in general relativity. As we will see, the values of the $w$ associated to the well-known particular solutions will provide a family of solutions depending on the parameter $\varkappa$ of the $f(R,T)$ gravity. We stress that particular choices of the parameter $w$ regarding the solution obtained in $f(R,T)$ do not necessarily have the same interpretation as the solutions obtained in general relativity with the same value of $w$.

\subsection{Black hole surrounded by a dust field}
\begin{figure}[H]
\centering
\includegraphics[scale=0.30]{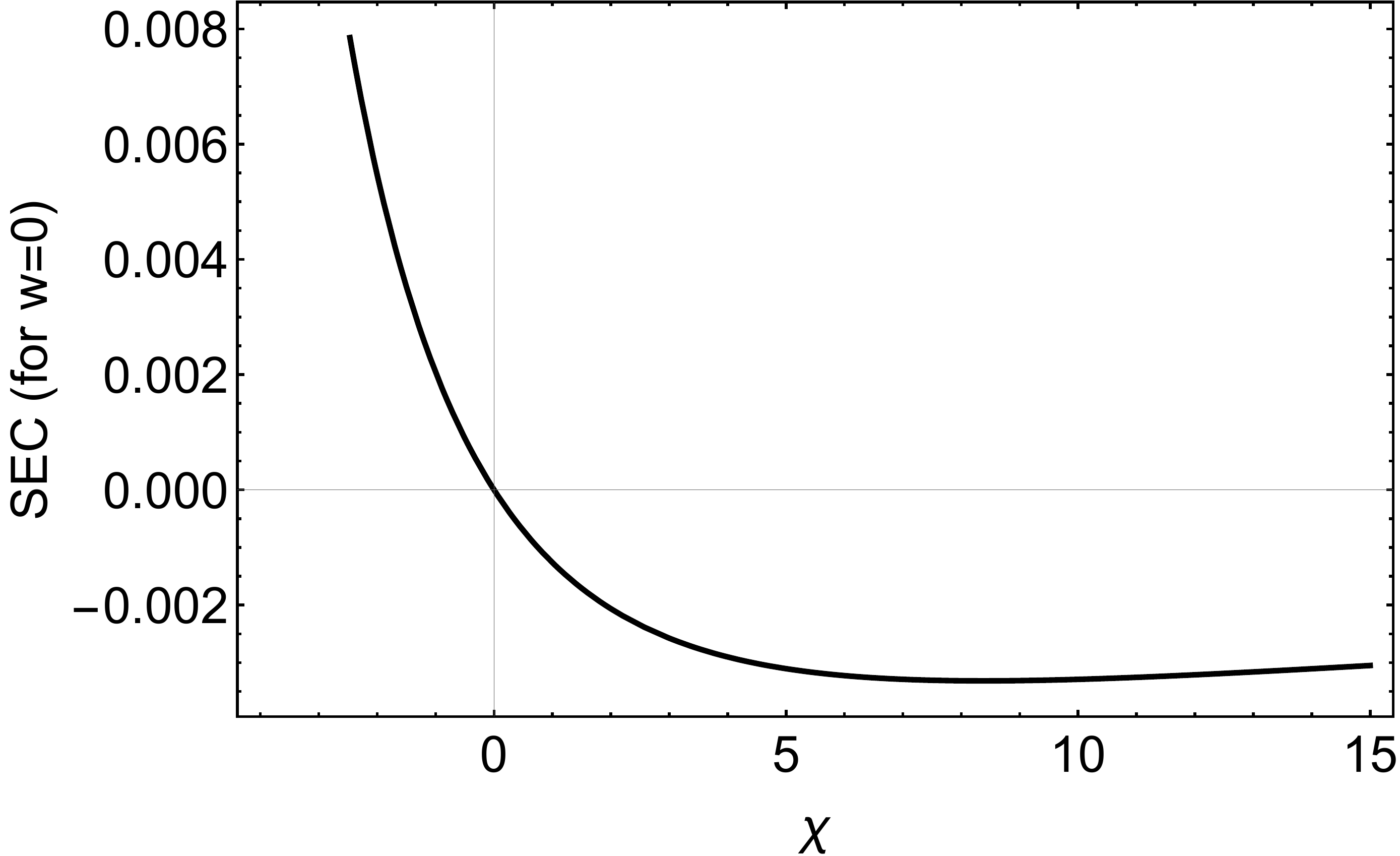} 
\caption{Condition (\ref{e22}) is plotted as a function of $\varkappa$ considering positive values of $K$. Negative values of this expression  are associated to the regions where the SEC is violated.   
}
\label{fig3}
\end{figure}

In this case we choice $w=0$ \cite{kiselev,Rastall} such that Eq.(\ref{eq14}) reduces to the form 
\begin{equation}
    B(r)=1-\frac{2M}{r}+Kr^{-\frac{8\pi}{3\varkappa+8\pi}}.
    \label{e20}
\end{equation}
The presence of the parameter $\varkappa$ imply that this solution in not equivalent to the metric of the black hole surrounded by a dust field in the context of general relativity. In the case of $\varkappa \rightarrow$ 0, Eq. (\ref{e20}) reduces to the metric associated to a Schwarzschild black hole with an effective mass $M_{eff}=2M-K$. The energy density associated to Eq. (\ref{e20}) is given by 
\begin{equation}
\rho=-\frac{3K \varkappa r^{-\frac{6(4\pi +\varkappa)}{8\pi+3\varkappa}}}{(8\pi+3\varkappa)^2},
    \label{e21}
\end{equation}
which in turn differs from the Kiselev black hole\cite{kiselev},in general relativity. The SEC conditions, represented by Equations (\ref{e19a}) and (\ref{e19b}), are equivalent in the case $w=0$ and are given by expression 
\begin{equation}
\frac{-K\varkappa}{64\pi^2+48\pi\varkappa+9\varkappa^2} \geq 0
    \label{e22}.
\end{equation} 
For $K>0$, the solution should satisfy the inequality $\varkappa \leq 0$ with $\varkappa \neq -8\pi/3$. In Fig. \ref{fig3}, the region where the SEC is satisfied, according conditions imposed by Eq.\ref{e22}, is plotted. In this way, in general relativity ($\varkappa=0$) the SEC is satisfied. In turn, in $f(R,T)$ gravity the SEC is satisfied or not, depending on values of $\varkappa$.

\begin{figure}[H]
\centering
\includegraphics[scale=0.21]{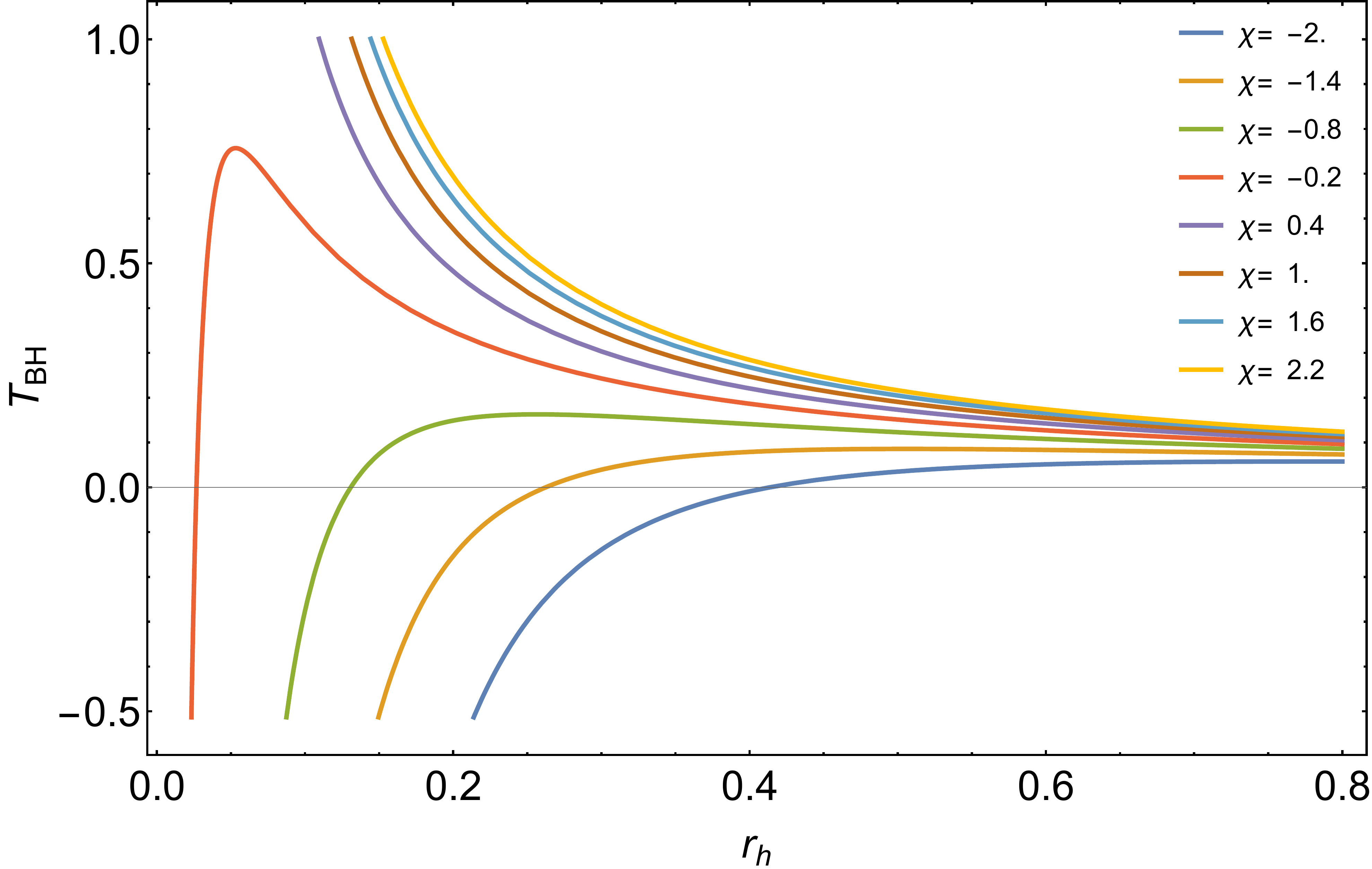} 
\caption{Equation (\ref{t4}) is plotted as a function of $r_h$ considering several values for $\varkappa$ with $K=\hbar =1$, and $w=0$. }
\label{fig3b}
\end{figure}

The Hawking temperature can be obtained by doing $w=0$ in Eq. (\ref{t3}), and is given by
\begin{equation}
T_{BH}=\frac{8\pi+3\varkappa (1+K r_h^{-\frac{8\pi}{8\pi+3\varkappa}})}{4\pi r_h(8\pi +3\varkappa)}.
    \label{t4}
\end{equation}
This equation gives us a family of curves depending on the values of $r_h$ and $\varkappa$. In Fig. \ref{fig3b}, we draw a set of graphs of the Hawking temperature $T_{BH}$, with respect to $r_h$ for different values of $\varkappa$. For positive values of $\varkappa$, the temperature remains positive, and increases when the value of $r_h$ decreases. In contrast, negative values of $\varkappa$ are associated to negative temperatures for small values of $r_h$.     

\subsection{Black hole surrounded by the radiation field}
We consider a black hole in which $w=1/3$, according to Ref. \cite{kiselev,Rastall} this specific value reproduces a black hole surrounded by the radiation field in the context of general relativity. By using the value $w=1/3$ in Eq. (\ref{eq14}) the metric function can be written as 
\begin{equation}
    B(r)=1-\frac{2M}{r}+Kr^{-\frac{3\pi}{\varkappa+3\pi}-1},
    \label{e23}
\end{equation}
due to the presence of the parameter $\varkappa$, the metric associated to this expression is different from the metric of a black hole surrounded by the radiation field in general relativity. In the limit of $\varkappa \rightarrow 0$, we obtain the metric function
\begin{equation}
B(r)=1-\frac{2M}{r}+\frac{K}{r^2},
    \label{e24}
\end{equation}
corresponding to the black hole surrounded by the radiation field obtained in \cite{kiselev}. The metric function (\ref{e24}) is effectively the metric of Reissner–Nordström black hole with an effective charge $Q_{eff}^2=K$.  
In this case, the energy density associated to solution (\ref{e23}) is written as 
\begin{equation}
\rho=\frac{9K\pi r^{-3\frac{4\pi+\varkappa}{3\pi+\varkappa}}}{8(3\pi+\varkappa)^2},
    \label{e25}
\end{equation}
that depends on the parameter of the $\varkappa$ of the $f(R,T)$ gravity, which makes this result different from the one  corresponding to the same system in general relativity, as expected. Regarding the SEC condition,  represented by Equations (\ref{e19a}) and (\ref{e19b}), these relations, in the case $w=1/3$, assume the form
\begin{align}
\frac{3\pi K}{4(3\pi+\varkappa)^2} \geq 0,\label{e26}\\
\frac{\pi K}{2(3\pi+\varkappa)^2} \geq 0.\label{e27}
\end{align}
If one consider $K \geq 0$, these expressions obtained imply that $\varkappa \neq -\pi/3$. 

\begin{figure}[H]
\centering
\includegraphics[scale=0.30]{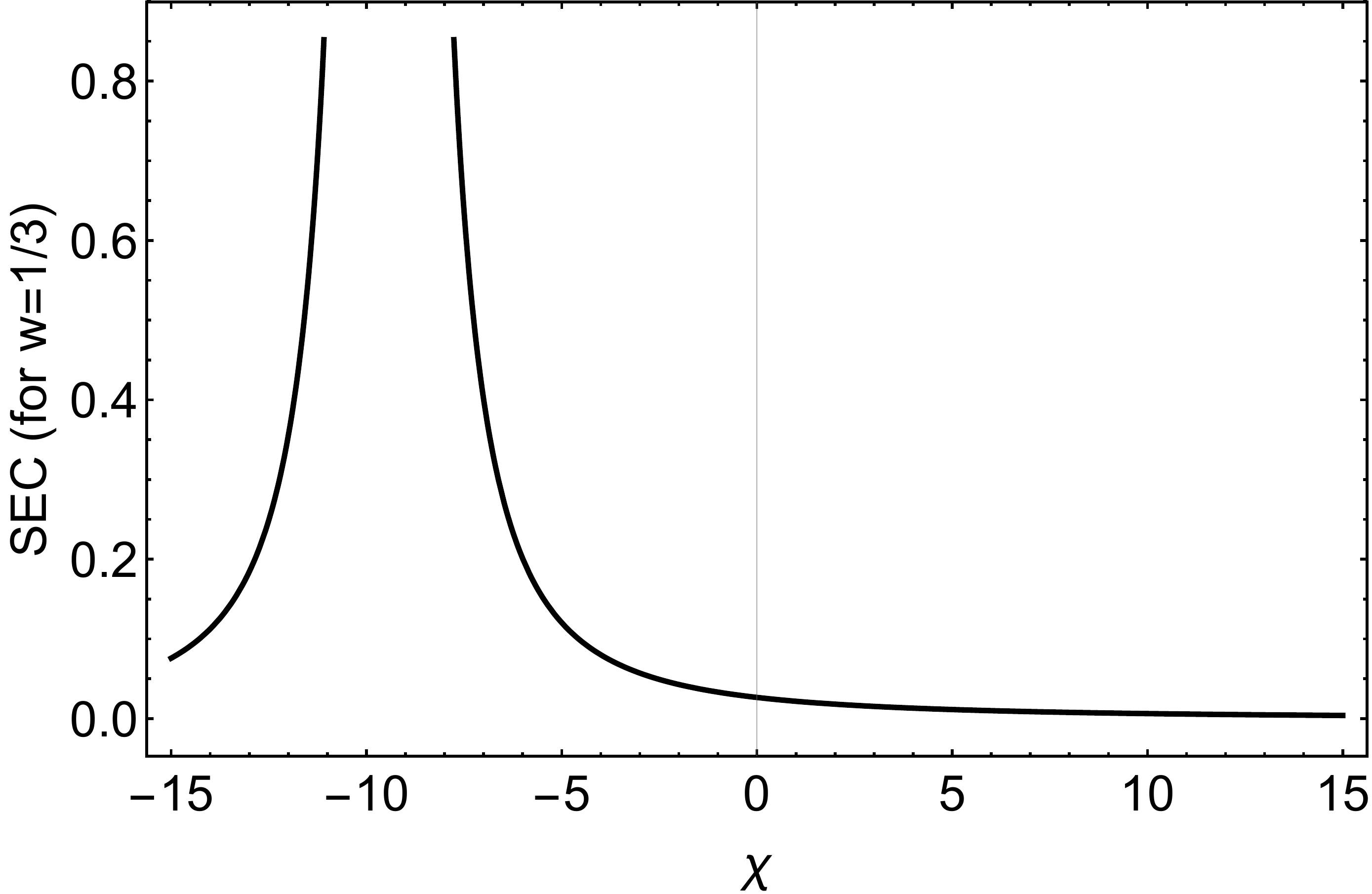} 
\caption{ Conditions given by Eqs. (\ref{e26}) and (\ref{e27}) are plotted as a function of $\varkappa$ considering positive values of $K$. Negative values of this expression  are associated to the regions where the SEC is violated.   
}
\label{fig4}
\end{figure}
\begin{figure}[H]
\centering
\includegraphics[scale=0.21]{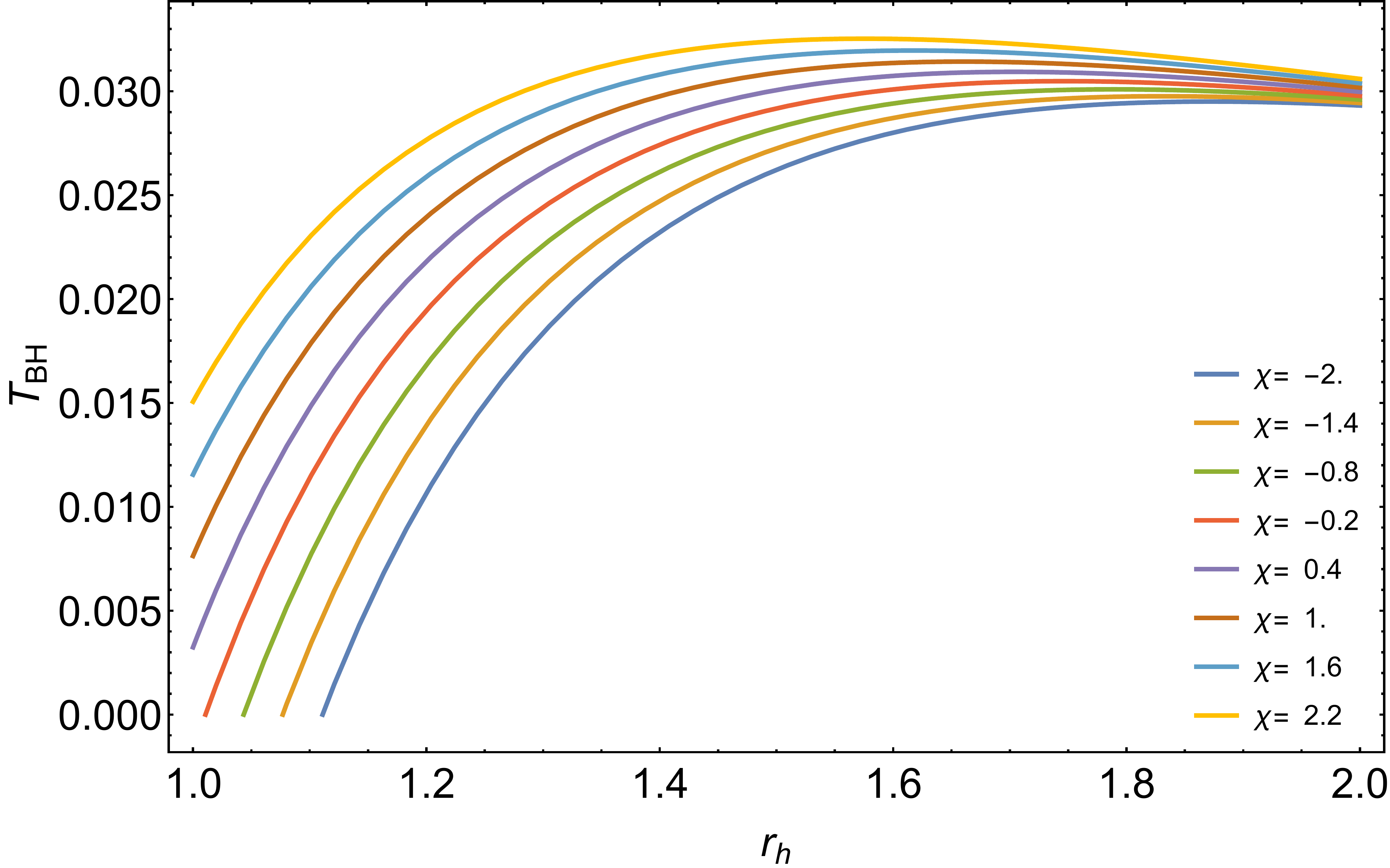} 
\caption{The Hawking temperature, given in Equation (\ref{t5}), is plotted as a function of $r_h$, considering several values for $\varkappa$ with $K=\hbar =1$, and $w=1/3$. }
\label{fig4b}
\end{figure}

In Fig. \ref{fig4}, the regions where the SEC is satisfied are plotted. Note that the SEC is fulfilled everywhere except at $\varkappa = -3\pi$ for $K>0$, where there is a divergence in the equations for the SEC. For $K<0$, the SEC is violated everywhere.  
By doing $w=0$ in Eq. (\ref{t3}), we obtain the Hawking temperature for the black hole surrounded by the radiation field, given by 
\begin{equation}
T_{BH}=\frac{1}{4\pi r_h}-\frac{3K r_h^{-2-\frac{3\pi}{3\pi+\varkappa}}}{4(3\pi+\varkappa)}.
    \label{t5}
\end{equation}
  We draw a set of graphs of the Hawking temperature $T_{BH}$ with respect to $r_h$ for different values of $\varkappa$ in Fig. \ref{fig4b}. For all values of $\varkappa$, the temperature remains positive in a region of considered and tends to negative values in the region where $r_h$ is small.     

\subsection{Black hole surrounded by the quintessence field}
\begin{figure}[H]
\centering
\includegraphics[scale=0.30]{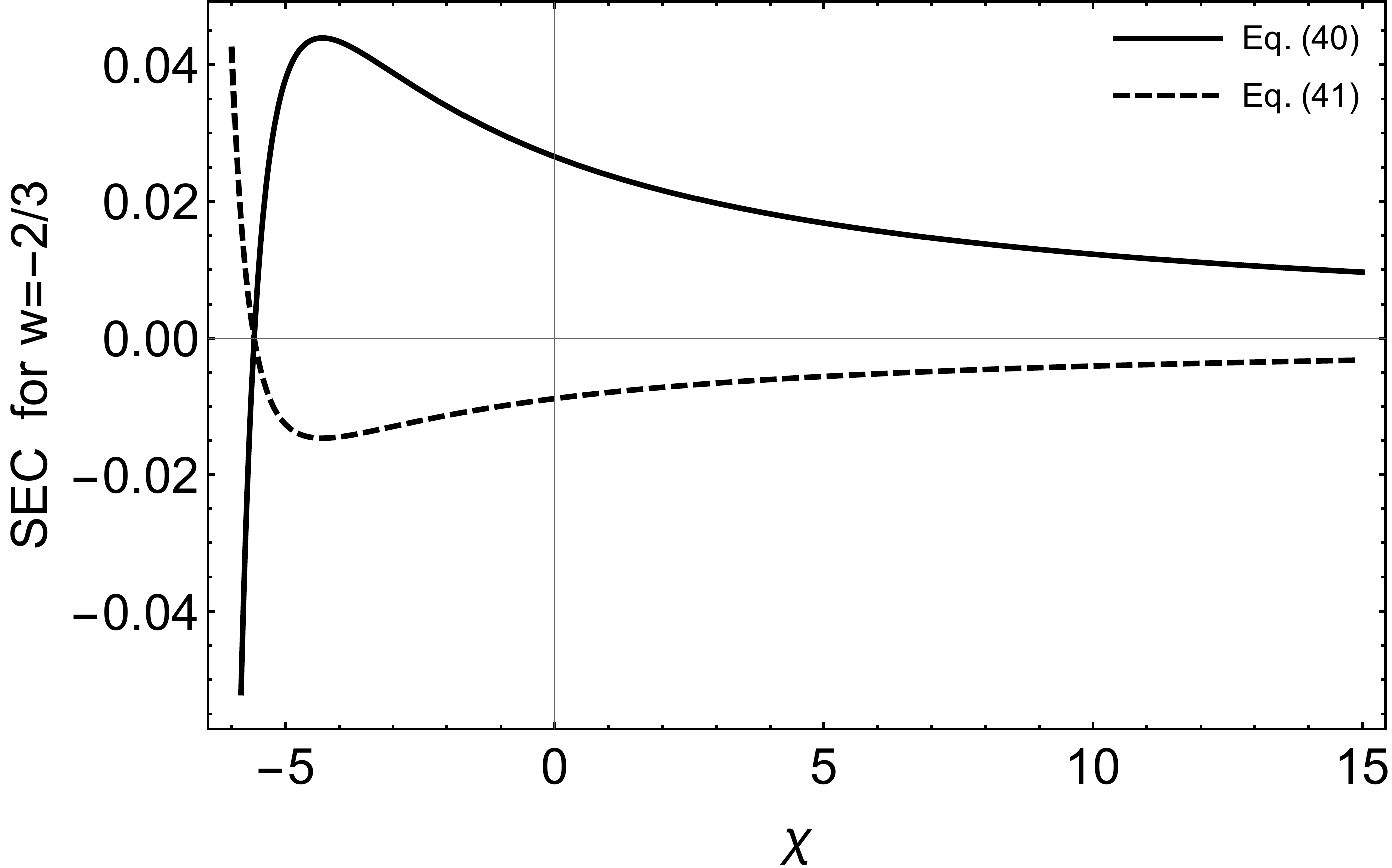} 
\caption{Conditions (\ref{e29}) and (\ref{e30}) originated in (\ref{e19a}) and (\ref{e19b})  plotted as a function of $\varkappa$ considering positive values of $K$. Negative values of ordinate axis are associated to the regions where the SEC is violated. There is one point where the SEC is satisfied, corresponding to the intersection of the curves.  
}
\label{fig5}
\end{figure}

\begin{figure}[H]
\centering
\includegraphics[scale=0.21]{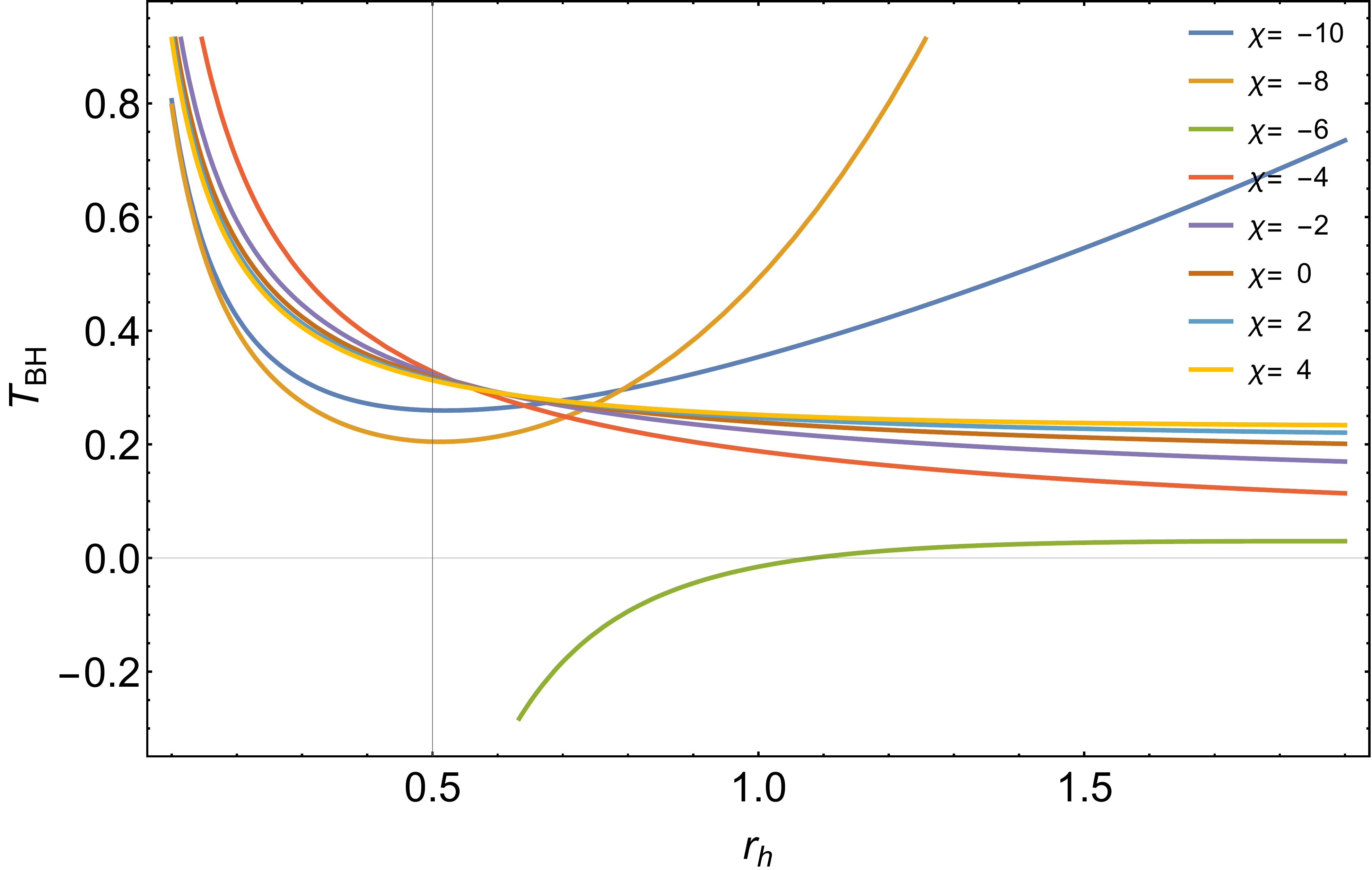} 
\caption{The Hawking temperature for the black hole surrounded by the quintessence field plotted as a function of $r_h$ considering different values of $\varkappa$ with $K=\hbar =1$, and $w=-2/3$. }
\label{fig5b}
\end{figure}

It is assumed that the parameter of the equation of state is $w=-2/3$ \cite{kiselev,Rastall} in this solution. Thus, the function $B(r)$ reads as
\begin{equation}
B(r)=1-\frac{2M}{r} + K r^{\frac{8(3\pi+2\varkappa)}{24\pi+11\varkappa}},
    \label{e28}
\end{equation}
we can see that the presence of term $\varkappa$, imply in a family of solutions associated to $w=-2/3$ in the context of $f(R,T)$ gravity. The SEC conditions, (\ref{e19a}) and (\ref{e19b}) for $w=-2/3$, take the following form 
\begin{align}
\frac{3K(16\pi +9\varkappa)}{(24\pi+11\varkappa)^2} \geq 0,\label{e29}\\
\frac{K(-16\pi-9\varkappa) }{(24\pi+11\varkappa)^2} \geq 0,\label{e30}  
\end{align}
once again, this expression depends on the signal of the constant $K$. But in this case, namely, $w=-2/3$, the condition in which the SEC is not violated is given by the pair of equations that have different behavior in the domain under consideration. Now, there is only a point where Equations (\ref{e29}) and (\ref{e30}) are satisfied, given by $\varkappa = -16\pi/9$. As we can see in Fig. \ref{fig5}, Eq. (\ref{e29}), represented by the continuous line and Eq. (\ref{e30}), represented by the dashed line, have opposite behavior. In this figure, it was considered positive values for $K$. By considering negative values, the shapes of curves are reversed. 
Assuming $w=-2/3$ in Eq. (\ref{t3}), the Hawking temperature for the black hole surrounded by the quintessence field takes the form
\begin{equation}
T_{BH}=\frac{3K (9 \varkappa +16 \pi ) r_{h}^{\frac{8 (2 \varkappa +3 \pi )}{11 \varkappa +24 \pi }}+11 \varkappa +24 \pi }{4 \pi  r_h (11 \varkappa +24 \pi )},
    \label{t6}
\end{equation}
 We can see in Fig. \ref{fig5b} the Hawking temperature $T_{BH}$ as a function of $r_h$ for different values of $\varkappa$ . In this case, the shape of curves changes significantly for each value of $\varkappa$. Note that all the curves are associated to the positive values of temperature, except the green curve, where the Hawking temperature assumes negative values in the domain studied.  
      
\subsection{Black hole surrounded by the cosmological constant field}
\begin{figure}[H]
\centering
\includegraphics[scale=0.30]{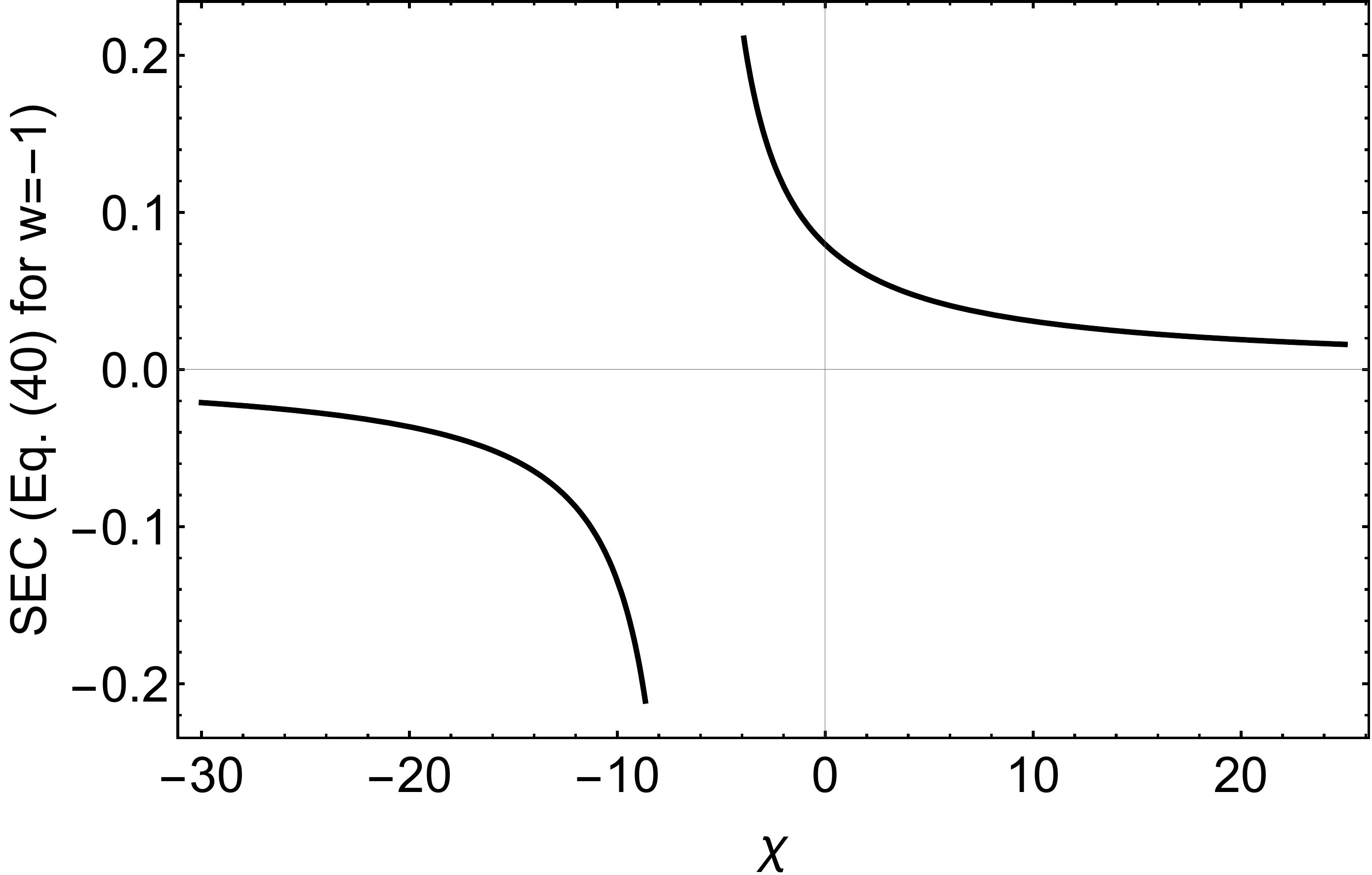} 
\caption{Condition (\ref{e32})  plotted as a function of $\varkappa$ considering positive values of $K$. Negative values of vertical axis are associated to the regions where the SEC is violated. In this graph, the SEC is not violated if $\varkappa >-2\pi$.  
}
\label{fig6}
\end{figure}

According to Ref. \cite{kiselev,rastall1}, the value $w=-1$ corresponds to the black hole surrounded by the cosmological constant field. 
Then, Eq. (\ref{eq14}), in the context of the $f(R,T)$ gravity, considering the equation of state with $w=-1$, reduces to
\begin{equation}
B(r)=1-\frac{2M}{r}+Kr^2.
    \label{e31}
\end{equation}
This is the same function obtained in \cite{kiselev} in the context of general relativity. Thus, we conclude that the case $w=-1$ in $f(R,T)$ gravity corresponds, exactly, to the one obtained in the framework of general relativity. The energy density assumes the form 
\begin{equation}
\rho = -\frac{3K}{8\pi + 4\varkappa}
    \label{e31b},
\end{equation}
where $\varkappa \neq -2\pi$ in order to avoid the singularity at this point.
Using $w=-1$ in Equations (\ref{e29}) and (\ref{e30}), we conclude that  Eq. (\ref{e30}) is zero and Eq. (\ref{e29}) can be written as
\begin{equation}    
\frac{K}{4\pi+2\varkappa} \geq 0,
\label{e32}
\end{equation}
with $\varkappa >-2\pi$. In Fig. \ref{fig6}, we show the behavior of the LHS of Eq. (\ref{e32}) considering $K>0$.

\subsection{Black hole surrounded by the phantom field}
\begin{figure}[H]
\centering
\includegraphics[scale=0.30]{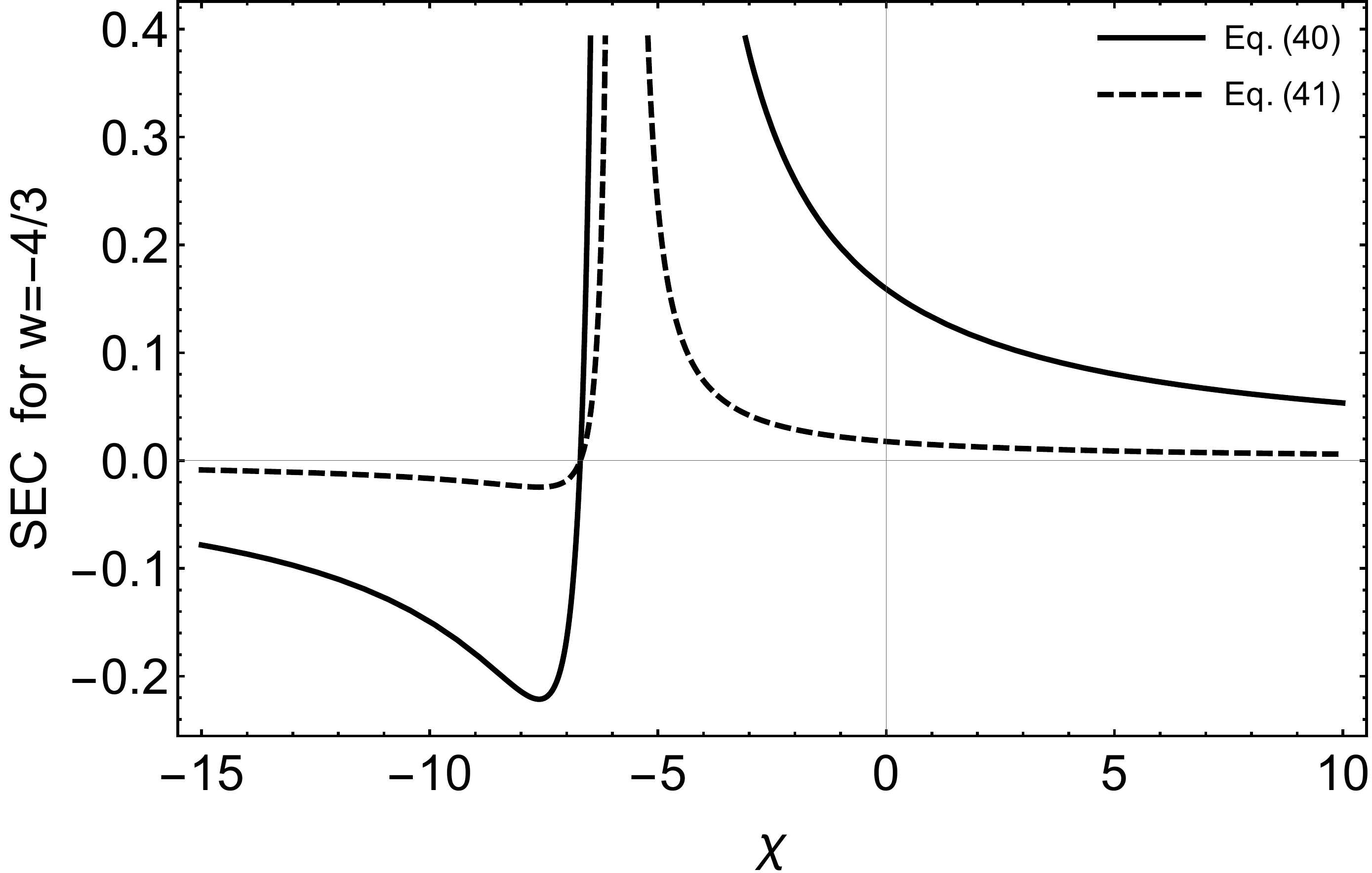} 
\caption{Conditions (\ref{e29}) and (\ref{e30}) originated from (\ref{e19a}) and (\ref{e19b})  plotted as a function of $\varkappa$ considering positive values of $K$. Negative values of ordinate axis are associated to the regions where the SEC is violated. There is one point where the SEC is satisfied, corresponding to the intersection of the curves.  
}
\label{fig7}
\end{figure}

The black hole associated to the phantom field can be obtained in general relativity, considering $w=-4/3$ \cite{rastall1}. Substituting this value in Eq. (\ref{eq14}), we find the following metric function
\begin{equation}
B(r)=1-\frac{2M}{r} + Kr^{\frac{8(9\pi+4\varkappa)}{24\pi+13\varkappa}}
    \label{e33}.
\end{equation}
The presence of the parameter $\varkappa$, imply that the solutions associated to $w=-4/3$ in the context of $f(R,T)$ gravity and the solutions with $w=-4/3$ in general relativity, are not equivalent. The SEC conditions, (\ref{e19a}) and (\ref{e19b}) for $w=-4/3$ are given by relations
\begin{align}
\frac{9K(32\pi +15\varkappa)}{(24\pi+13\varkappa)^2} \geq 0,\label{e34}\\
\frac{K(32\pi +15\varkappa) }{(24\pi+13\varkappa)^2} \geq 0,\label{e35}  
\end{align}
 where the solution of the    pair of equations above demands that
 \begin{equation}
 -\frac{32\pi}{15} \leq \varkappa < -\frac{24\pi}{13},\:\:\text{or}\:\:\:\: \varkappa >-\frac{24\pi}{13}.
     \label{e36}
 \end{equation}
   In Fig. \ref{fig7}, Eqs. (\ref{e35}) and (\ref{e36}), represented by the continuous line and dashed lines, respectively, have similar behavior. In this figure, it was considered positive values for $K$. The shapes of the curve for negative values of $K$, are reversed as compared to the one for positive values of $K$. 

\begin{figure}[H]
\centering
\includegraphics[scale=0.21]{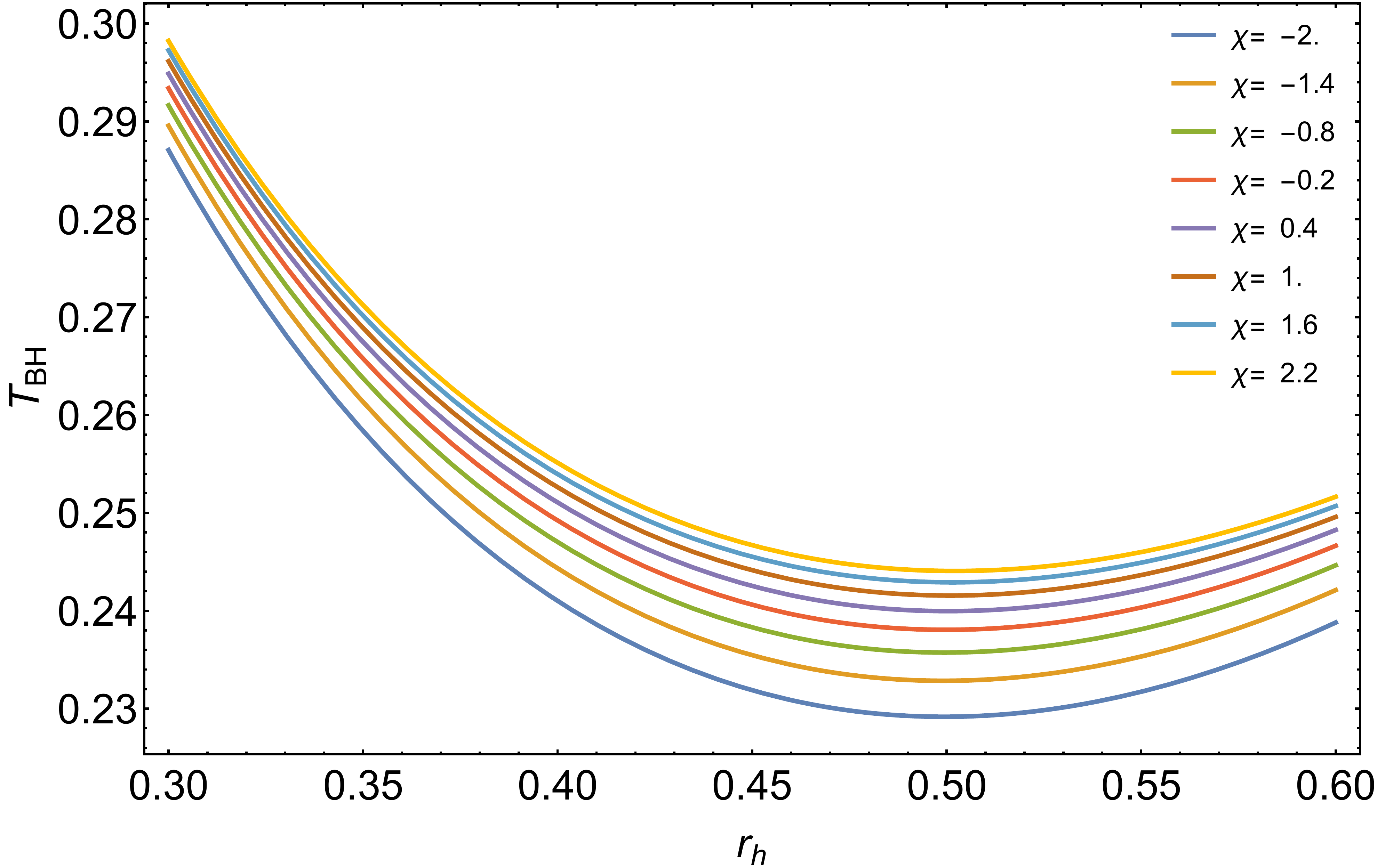} 
\caption{The Hawking temperature for the black hole surrounded by the quintessence field plotted as a function of $r_h$ considering different values of $\varkappa$ with $K=\hbar =1$, and $w=-4/3$. }
\label{fig6b}
\end{figure}

Assuming $w=-4/3$ in Eq. (\ref{t3}), the Hawking temperature for the black hole surrounded by the quintessence field takes the form
\begin{equation}
T_{BH}=\frac{3 K (15 \chi +32 \pi ) r^{\frac{8 (4 \chi +9 \pi )}{13 \chi +24 \pi }}+13 \chi +24 \pi }{4 \pi  r (13 \chi +24 \pi )}.
    \label{t7}
\end{equation}
   The Hawking temperature, in this case, is shown in Fig. \ref{fig6b}. The shape of curves do not change significantly for each value of $\varkappa$. We observe that all the curves are associated to the positive values of temperature in the domain studied.  
      
\section{Final remarks}
The results obtained in this paper provide, for the first time in the literature, a solution of the gravitational field equation in $f(R,T)$ gravity corresponding to the fluid of Kiselev's. This solution has an interesting feature: by choosing particular values of the parameter of the equations of state, namely $w$, we are able to reproduce well-known solutions of the Einstein field equation as particular cases. As we have seen in this paper, this remarkable feature is present in the context of $f(R,T)$ gravity.

This fluid has been studied in the context of general relativity and modified theories of gravity. In the present paper, we have considered the model $f(T)=\varkappa T^{n}$ and the conditions that it must satisfy in order to generate Kiselev black holes in the context of this theory of gravity. Indeed, the additivity and linearity conditions suggests that
the accepted value of $n$ should be $1$. The general solution obtained  has an additional structure that comes from the dependence on the parameter $\varkappa$ of the $f(R,T)$ gravity. This property implies that several particular values of the parameter $w$ in modified gravity lead to solutions which differs from the Kiselev black hole, in general relativity. 

To carry out a systematic analysis of the solution obtained in $f(R,T)$ gravity, we use particular  values for $w$ associated to the solutions corresponding to a black holes surrounded by dust field ($w=0$), radiation field ($w=1/3$), quintessence field ($w=-2/3$), cosmological constant field ($w=-1$), phantom field ($w=-4/3$) and so on. Considering the particular solutions studied, only the case $w=-1$ is equivalent to the solution obtained in general relativity. The other cases studied, have an additional structure in the solutions provided by the modified gravity, which is characterized by dependence on the parameter $\varkappa$.

Due to the presence of the additional structure from $f(R,T)$ gravity, in the majority of the solutions considered, the SEC condition can be satisfied, considering the particular cases. We have analyzed in details the conditions imposed by SEC on the parameter $\varkappa$ of the theory and on the constant $K$ and conclude that the particular solutions that depend on $\varkappa$ are more flexible regarding the energy conditions. We studied the horizons associated to the solution obtained and determined the Hawking temperatures. As we can see, in some particular cases studied, the Hawking temperature can reach negative values for certain values of $\varkappa$, which means that some restrictions should be imposed to the values of $\varkappa$ in order to avoid this behavior.  

Finally, we observe that the choice of models of $f(R,T)$ gravity with high order terms in $R$ or $T$ can lead to difficulties in finding exact solutions due to the condition $H=kF$, supposed in the solution obtained.

\acknowledgments
LCNS would like to thank Conselho Nacional de Desenvolvimento Cient\'ifico e Tecnol\'ogico (CNPq) for partial financial support through the research Project No. 164762/2020-5. I. P. L. was partially supported by the National Council for Scientific and Technological Development - CNPq grant 306414/2020-1 and by the grant 3197/2021, Para\'iba State Research Foundation (FAPESQ). I. P. L. would like to acknowledge the contribution of the COST Action CA18108. V.B.B. is partially supported by CNPq through the Research Project No. 307211/2020-7.

\bibliography{referencias_unificadas}

\end{document}